\documentclass{aastex701}
\usepackage{amsmath}
\usepackage[utf8]{inputenc} 
\usepackage{CJK}

\begin{document}

\title{Measuring the Coronal Magnetic Field with 2D Coronal Seismology: A Forward-Modeling Validation}

\begin{CJK*}{UTF8}{gbsn}
\author[orcid=0000-0002-4973-0018]{Zihao Yang}

\affiliation{High Altitude Observatory, National Center for Atmospheric Research, Boulder, CO, USA, 80301}
\email[show]{zihao@ucar.edu}  

\author[orcid=0000-0001-5850-3119]{Matthias Rempel} 
\affiliation{High Altitude Observatory, National Center for Atmospheric Research, Boulder, CO, USA, 80301}
\email{rempel@ucar.edu}

\author[orcid=0000-0001-9831-2640]{Sarah Gibson}
\affiliation{High Altitude Observatory, National Center for Atmospheric Research, Boulder, CO, USA, 80301}
\email{sgibson@ucar.edu}

\author[orcid=0000-0002-8439-6166]{Giuliana de Toma}
\affiliation{High Altitude Observatory, National Center for Atmospheric Research, Boulder, CO, USA, 80301}
\email{detoma@ucar.edu}

\begin{abstract}

In recent years, a two-dimensional (2D) coronal seismology technique applied to spectral-imaging data from the Coronal Multi-channel Polarimeter (CoMP) and UCoMP has enabled routine measurement of the global  coronal magnetic field. The technique combines coronal transverse wave phase speed from Doppler measurements with electron densities from the Fe \sc{xiii}\rm{} 10798/10747 \AA\ intensity ratio to infer the magnetic field strength, while the wave propagation directions from Doppler measurements trace the magnetic field direction. To validate the accuracy and robustness of this method, we use forward modeling of a MURaM simulation that produces open and closed magnetic structures with excited waves. From the synthetic Doppler velocity, Fe \sc{xiii}\rm{} infrared line intensities, and linear polarization signals, we apply the 2D coronal seismology technique to estimate the magnetic field strength and direction. A comparison with the simulation ground truth shows close agreement, indicating that the technique can recover the line-of-sight emissivity-weighted magnetic field direction and strength with high accuracy. We also perform a parameter-space analysis to quantify sensitivities of the method to parameter choice. These findings provide practical guidance for CoMP/UCoMP-like analysis and demonstrate that 2D coronal seismology can deliver reliable, LOS emissivity-weighted measurements of the coronal magnetic field from coronal wave observations.

\end{abstract}


\section{Introduction} 

The Sun is a magnetized star, and its magnetic field links the solar interior to its outer atmosphere and the expanded heliosphere. The magnetic field shapes the structures and dynamics of the solar atmosphere, powering energetic events such as solar flares and coronal mass ejections, contributing to the plasma heating, and modulating the solar activity cycles. The central role of solar magnetic field makes it one of the most essential parameters to measure in solar physics. At the photospheric level, the magnetic field can now be measured nearly continuously, offering uninterrupted daily coverage. By contrast, routine measurements of the upper atmosphere, especially the solar corona, have long remained elusive.

In the absence of routine coronal magnetic field measurements, information about the coronal magnetic field is often obtained indirectly through extrapolations from routinely measured photospheric magnetograms. These approaches range from force-free extrapolations \citep[e.g.,][]{2012LRSP....9....5W} to more advanced magnetohydrostatic \citep[e.g.,][]{2018ApJ...866..130Z} and magnetohydrodynamic models \citep[e.g.,][]{2009ApJ...690..902L,2022Innov...300236J}, which attempt to incorporate plasma forces and dynamic effects. Despite their widespread use, such models usually rely on some assumptions and photospheric boundary conditions that are not always valid in the solar corona. Over the past several decades, a variety of attempts have been pursued to measure the coronal magnetic field through coronal observations. One approach extends the spectro-polarimetric measurements from the photospheric magnetic field to the coronal one. Using the spectro-polarimetric observations from Fe \sc{xiii}\rm{} 10747 \AA\ line, \cite{2000ApJ...541L..83L, 2004ApJ...613L.177L} derived the line-of-sight (LOS) magnetic field strength in the corona above active regions. Such measurements generally require large-aperture telescopes and deep exposures. With the deployment of the largest solar telescope, Daniel K. Inouye Solar Telescope \citep[DKIST,][]{2020SoPh..295..172R}, high-resolution coronal spectropolarimetry has become possible. Using the observations from DKIST/CryoNIRSP, \cite{2024SciA...10.1604S} obtained an off-limb coronal magnetogram above active regions, demonstrating the capability of DKIST in measuring coronal magnetic field based on spectro-polarimetric observations. Radio diagnostics provide a complementary pathway because coronal radio emissions are related to magnetic field strength, therefore, microwave imaging observations can be used to estimate the distribution of coronal magnetic field strength \citep[e.g.,][]{2020Sci...367..278F, 2020NatAs...4.1140C, 2021ApJ...923..213W}. These measurements are typically confined to regions with very strong field (e.g., strong active regions and flare regions). Another widely used approach to estimate the coronal magnetic field is the coronal seismology technique: combining coronal wave/oscillation observations and magnetohydrodynamic (MHD) wave theory allows for inference of coronal magnetic field strength from wave dispersion relations \citep[e.g.,][]{2001A&A...372L..53N, 2024A&A...681L...4G}. However, such measurements are usually applied to occasional events and can only provide a single value of the magnetic field strength in the target regions. None of the above methods has ever achieved routine and continuous measurements of the coronal magnetic field over large field-of-views (FOVs). Such routine, large-FOV measurements are required to capture the temporal evolution of the global coronal magnetic field and structure, providing the broader context needed for systematic investigations of coronal dynamic events, as well as for heliospheric modeling and space-weather forecasting.

Recently, based on the coronagraphic spectral imaging observations from the Coronal Multi-channel Polarimeter \citep[CoMP,][]{2008SoPh..247..411T} and its upgraded version, the Upgraded Coronal Multi-channel Polarimeter \citep[UCoMP,][]{2016JGRA..121.8237L}, which provides a larger FOV, enhanced signal-to-noise ratio, improved spatial resolution, and more stable data quality compared to CoMP, a new two-dimensional (2D) coronal seismology technique based on the traditional coronal seismology has been developed for mapping the global coronal magnetic field. It has been shown that the propagating Doppler velocity fluctuations observed by CoMP and UCoMP are signatures of propagating transverse waves in the corona \citep[e.g.,][]{2007Sci...317.1192T, 2009ApJ...697.1384T, 2015NatCo...6.7813M}. Under coronal conditions, their phase speed relates coronal magnetic field strength and coronal density \citep[][]{2017A&A...603A.101L, 2020Sci...369..694Y, 2024Sci...386...76Y}. The phase speed of the waves are obtained via a wave-tracking procedure on the Doppler velocity data \citep[][]{2009ApJ...697.1384T, 2015NatCo...6.7813M, 2020Sci...369..694Y, 2024Sci...386...76Y}, while the density is estimated through Fe \sc{xiii}\rm{} infrared spectral lines 10747/10798 \AA\ intensity ratio \citep[e.g.,][]{2020Sci...369..694Y, 2021ApJ...906..118D, 2024ApJ...966..122Z, 2024Sci...386...76Y}. Based on CoMP observations, combining the calculated phase speed from wave-tracking procedure and the density estimated from line ratio method, \cite{2020Sci...369..694Y} obtained the first map of the global coronal plane-of-sky magnetic field strength. Moreover, because the transverse waves propagate along field lines, the estimated wave propagation direction also recovers the information of coronal plane-of-sky magnetic field direction \citep[][]{2020ScChE..63.2357Y,2020Sci...369..694Y}. With the upgrade of UCoMP instrument, \cite{2024Sci...386...76Y} applied this technique to the continuous UCoMP observations, and generated 114 global maps of coronal magnetic field strength and direction across several solar rotations, providing the first routine measurements of the global off-limb coronal magnetic field. A comparison between measurements and coronal MHD models generated using the magnetohydrodynamics algorithm outside a sphere \citep[MAS,][]{2009ApJ...690..902L} code also demonstrated that the measured magnetic field is a LOS emissivity-weighted average of the plane-of-sky (POS) component of the coronal magnetic field \cite{2024Sci...386...76Y}.

Several previous studies have assessed the feasibility and accuracy of this technique using forward-modeling experiments. For example, \cite{2018ApJ...856..144M} synthesized Fe \sc{ix}\rm{} 171 \AA\ emission and transverse waves signatures from an ideal MHD model with a uniform 5 G field along a fixed direction. They manually added wave drivers at the lower boundary to obtain the Doppler velocity of the waves. \cite{2025RAA....25a5010G} performed a similar study, using another 3D MHD simulation to synthesize similar wave signatures in an open magnetic flux tube with gravitational stratification. They also used an initial magnetic field strength of 4 G and manually imposed wave drivers at the boundary of the simulation domain. Both works used the 2D coronal seismology technique to recover the input field strength with good accuracy. These tests, however, relied on simplified configurations such as open flux tubes with a fixed input magnetic field value. The simulated wave perturbations are driven by artificial boundary drivers that excite such oscillations, and are largely propagating in a fixed direction. Therefore, these tests generally validate idealized and simplified scenarios rather than diverse coronal conditions.

In this work, we use a realistic coronal MHD simulation that includes both open and closed field regions to perform the validation of the 2D coronal seismology technique through forward modeling. The wave signatures are spontaneously excited by the convective motions in the simulation domain without imposed wave drivers. Using synthesized observables from the simulation, we obtain the magnetic field directions and strengths, and compare them with the simulation ground truth. We also conduct a parameter-space analysis to identify the range of parameter settings during wave-tracking procedure that ensure robust and reliable diagnostics. This analysis provides quantitative guidance for application to real coronal observations.

\section{MHD Model and Observables Synthesis} 
\subsection{Fe \sc{xiii}\it{} Lines Used for 2D Coronal Seismology}\label{sec:fexiii}

Previous studies show that the Doppler velocity fluctuations seen by CoMP and UCoMP are signatures of kink waves \citep[e.g.,][]{2015NatCo...6.7813M,2019NatAs...3..223M}. The phase speed of kink waves is related to the magnetic field strength and density inside and outside of a flux tube:
\begin{equation}
    v_k=\sqrt{\frac{B^2_\text{i}+B^2_\text{o}}{\mu(\rho_\text{i}+\rho_\text{o})}}
\end{equation}
where the subscripts i and o refer to the physical quantities inside and outside of a flux tube, respectively, and $\mu$ is the magnetic permeability in the vacuum. In the corona, where the plasma-$\beta$ is typically much less than unity, magnetic pressure dominates. Under such low-$\beta$ conditions, pressure balance between the inside and outside of flux tubes requires that $B_\text{i}\sim B_\text{o}$ \citep[e.g.,][]{2009SSRv..149..199R, 2015NatCo...6.7813M, 2017A&A...603A.101L}. Furthermore, because the spatial resolution of CoMP and UCoMP is generally insufficient to resolve individual flux tubes \citep[][]{2013A&A...556A.104P}, the observed density can be regarded as an average over multiple unresolved flux tubes within each pixel ($\left<\rho\right>$). With these assumptions, the expression for the kink wave phase speed simplifies to
\begin{equation}\label{eq:vph}
    v_k=\frac{B}{\sqrt{\mu\left<\rho\right>}}
\end{equation} 

Operationally, the 2D coronal seismology technique consists of two main procedures: the wave tracking procedure, which provides $v_k$ in Eq. \ref{eq:vph}, requiring Doppler velocity time series from Fe \sc{xiii}\rm{} 10747 \AA\ line; and the density diagnostics procedure, which provides $\rho$, requiring the intensities of both Fe \sc{xiii}\rm{} 10747 \AA\ and 10798 \AA\ lines. Combining the phase speed with the density gives the magnetic field strength. To compare the measured magnetic field direction from the technique with magnetic azimuth derived using linear polarization, Stokes-Q and Stokes-U signals of Fe \sc{xiii}\rm{} 10747 \AA\ are also required.

\subsection{MHD Model}\label{sec:model}

\begin{figure}
    \centering
    \includegraphics[width=0.5\linewidth]{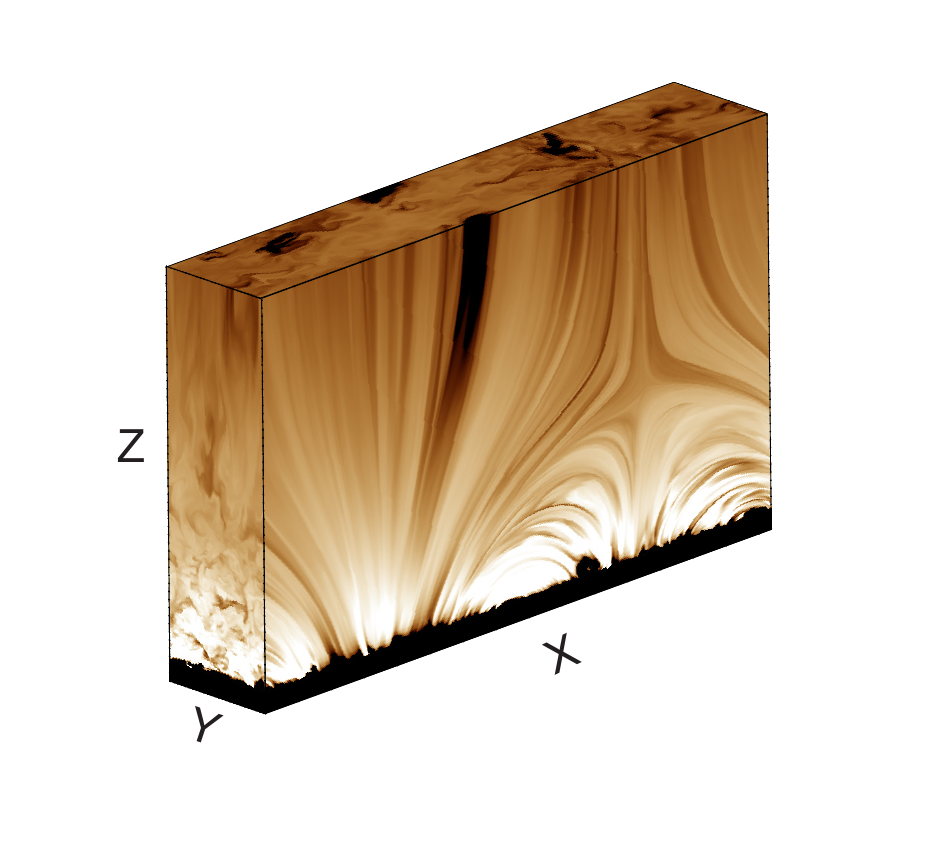}
    \caption{Three-dimensional visualization of the simulated coronal emissivity of Fe \sc{xiii}\rm{} from the MURaM simulation. The three orthogonal planes show emissivity cross sections at the selected slices, revealing the structures of the 3D data cube. The X, Y and Z axes correspond to the horizontal and vertical directions in the simulation domain.}
    \label{fig:3dcube}
\end{figure}

To synthesize the required coronal observables as described in Sect.\ref{sec:fexiii}, we use the coronal extension of the MURaM code \citep[][]{2017ApJ...834...10R} to simulate a domain that encompasses the upper convection zone, photosphere, chromosphere and lower solar corona in a domain with an extent of $294.912\times 49.152\times 196.608\ \text{Mm}^3$ in the x, y (horizontal) and height (vertical) directions, respectively. The numerical grid spacing is $384\times 384\times 96$ km. The position of the photosphere is about 8 Mm above the bottom boundary. Horizontal boundaries are periodic, while the top boundary imposes a potential magnetic field and is open for outflows. The bottom boundary, located beneath the convection zone, allows for convective energy flux to replenish energy that is radiating away in the photosphere
\citep[see,][for further detail]{2014ApJ...789..132R}. The simulation domain was relaxed in two stages. We first started with the lowermost 9.216 Mm of the domain (only convection zone/photosphere), in which we added an isentropic stratification and a magnetic field of the form 
\begin{equation}
    B_\text{z}=5+50\sin{\left(\dfrac{2\pi x}{294.912\ \text{Mm}}\right)}\ \text{G}
\end{equation}
Radiative cooling starts convection and the resulting convective flows will organize the initial magnetic field into magnetic field concentrations reminiscent of network field in the solar atmosphere reaching field strengths of up to 1-2 kG. We ran this stage for a total of 63 simulated solar hours. In the second stage we added the corona. To this end we filled the coronal part of the domain with a potential field extrapolation and a hydrostatic atmosphere with initially 500,000 K temperature in the corona (6,000 K for the chromospheric part in-between). The domain was then evolved for another 9 hours to allow for the corona to become dynamically relaxed. We note that the grid resolution of $384\times 384\times 96$ km is insufficient to resolve solar granulation and as a consequence convection cells are broadened to about twice the size of granules. Nonetheless, this convection pattern does excite waves and generates a Poynting flux that is sufficient to maintain the overlying corona. The coronal radiative EUV flux losses correspond to about $7\times 10^5\ \text{erg}\ \text{cm}^{-2}\ \text{s}^{-1}$, which is a few times stronger than typical quiet Sun losses \citep[][]{1977ARA&A..15..363W}. Figure \ref{fig:3dcube} shows a three-dimensional visualization of the emissivity slices through the simulation domain. In this study, we selected a specific viewing angle so that the majority of the coronal structures are aligned with the plane-of-sky (the XZ plane).

The magnetic field distribution we chose for this simulation leads to a coronal magnetic field that combines elements of open and closed field, with a pseudo streamer topology that has a null point at around 80 Mm height above the photosphere. This choice leads to a strong variation of Alfv\'{e}n velocity in the corona and a combination of relatively straight and strongly curved field lines, which allows thorough testing of the wave tracking approach under a diverse set of coronal conditions.

\subsection{Synthesis of Coronal Observables}

Based on the distributions of temperature, density, velocity and magnetic field from the simulation described in Sect. \ref{sec:model}, we synthesized the coronal line intensities, Doppler velocities and linear polarization signals of the two Fe \sc{xiii}\rm{} infrared lines. All forward modeling is carried out at the original spatial resolution of the simulation without additional spatial degradation.

The intensities of Fe \sc{xiii}\rm{} 10747 \AA\ and 10798 \AA\ are calculated as 
\begin{equation}\label{eq:intensity}
    I=\int{G(T, N_e, h)\cdot N_eN_H\ \text{d}l}
\end{equation}
where $G(T, N_e, h)$ (in unit of $\text{erg}\cdot \text{cm}^{3}\cdot \text{s}^{-1}\cdot\text{sr}^{-1}$) is the contribution function of the spectral line as a function of temperature ($T$), electron density ($N_e$) and coronal height ($h$); $N_H$ is the number density of Hydrogen; and $l$ is the distance along the integral direction \citep[e.g.,][]{2021ApJ...906..118D}. The intensities of these coronal infrared forbidden lines are impacted by both electron collisional excitation and photo-excitation, with the latter being dependent on coronal height. The effect of photo-excitation has been included in the contribution function $G(T, N_e, h)$. Under typical coronal temperature, hydrogen and helium are generally regarded as fully ionized. If we assume only contributions from the two most abundant elements (H, He) to electron density, and adopt the relative abundance of Helium as $N_H/N(\text{He})=10$ \citep[e.g.,][]{1985ApJS...57..173M}, we have $N_H/N_e\approx0.83$, thus Eq. \ref{eq:intensity} will be 
\begin{equation}\label{eq:intensity0.83}
    I=0.83\int{G(T, N_e, h)\cdot N_e^2\ \text{d}l}
\end{equation}

To efficiently calculate the synthesized intensities of the two target lines, we created three-dimensional look-up tables of the contribution functions of the two lines $G(T, N_e, h)$ over temperature, electron density and height (which accounts for the height-dependent photo-excitation) using the routine \textit{gofnt.pro} from the \textsc{chianti} version 11 database \citep[][]{1997A&AS..125..149D,2024ApJ...974...71D}. For each grid point in the simulation, $G$ was interpolated at the local $(T, N_e, h)$ to compute the volume emissivity at that point as $\varepsilon=N_e^2G$. The LOS-integrated intensity was then obtained as $I(l)=0.83\int{N_e^2(i)G(i)\text{d}l}$, where $l$ is the LOS path and $i$ is each point position along the LOS. 

To simulate the Doppler velocity data as measured by CoMP and UCoMP, we computed the emissivity-weighted LOS average Doppler velocity using
\begin{equation}
    v_{\text{LOS}}=\frac{\int{v_{\text{LOS, }i}\cdot\varepsilon_{i}\ \text{d}l}}{\int{\varepsilon_{i}\ \text{d}l}}
\end{equation}
where $\varepsilon_{i}$ is the emissivity calculated using the above method, and $v_{\text{LOS, }i}$ is the LOS Doppler velocity at location $i$ along the LOS.

We also synthesized the LOS-integrated linear polarization signals (Stokes Q and U) using the \textsc{forward} package \citep[][]{2016FrASS...3....8G}. The magnetic azimuth was then calculated as 
\begin{equation}\label{eq:azimuth}
    \phi=\frac{1}{2}\tan^{-1}{\frac{U}{Q}}
\end{equation}.

\section{Validation of Two-dimensional Coronal Seismology} \label{sec:floats}

The 2D coronal seismology technique has been applied to CoMP and UCoMP data to derive the coronal magnetic field. The technique has been described in detail in previous literature \citep[e.g.,][]{2020ScChE..63.2357Y,2020Sci...369..694Y, 2024Sci...386...76Y, 2025RAA....25a5010G}, and we will briefly recap the core workflow of this technique here. 

As shown in Eq. \ref{eq:vph}, two main steps are required to obtain the coronal magnetic field: density diagnostics and wave tracking. The latter will provide both wave propagation direction and wave phase speed. As discussed in previous literature, the wave propagation direction also indicates the magnetic field direction on the plane-of-sky. Therefore, through these two steps, we are able to obtain both the magnetic field direction and strength.

\subsection{Density Diagnostics}
Atomic physics shows that the intensity ratio of the two Fe \sc{xiii}\rm{} infrared lines provides a diagnostic of the local electron density. We first computed the theoretical relationship between electron density and the intensity ratio of Fe \sc{xiii}\rm{} 10798/10747 \AA\ from \textsc{chianti} version 11 database. Using the synthetic Fe \sc{xiii}\rm{} 10798 \AA\ and 10747 \AA\ intensities from the simulation, we could calculate the intensity ratio at each pixel and use the theoretical relationship to obtain the electron density, yielding a 2D map of the diagnosed density. 

\begin{figure}
    \centering
    \includegraphics[width=0.95\linewidth]{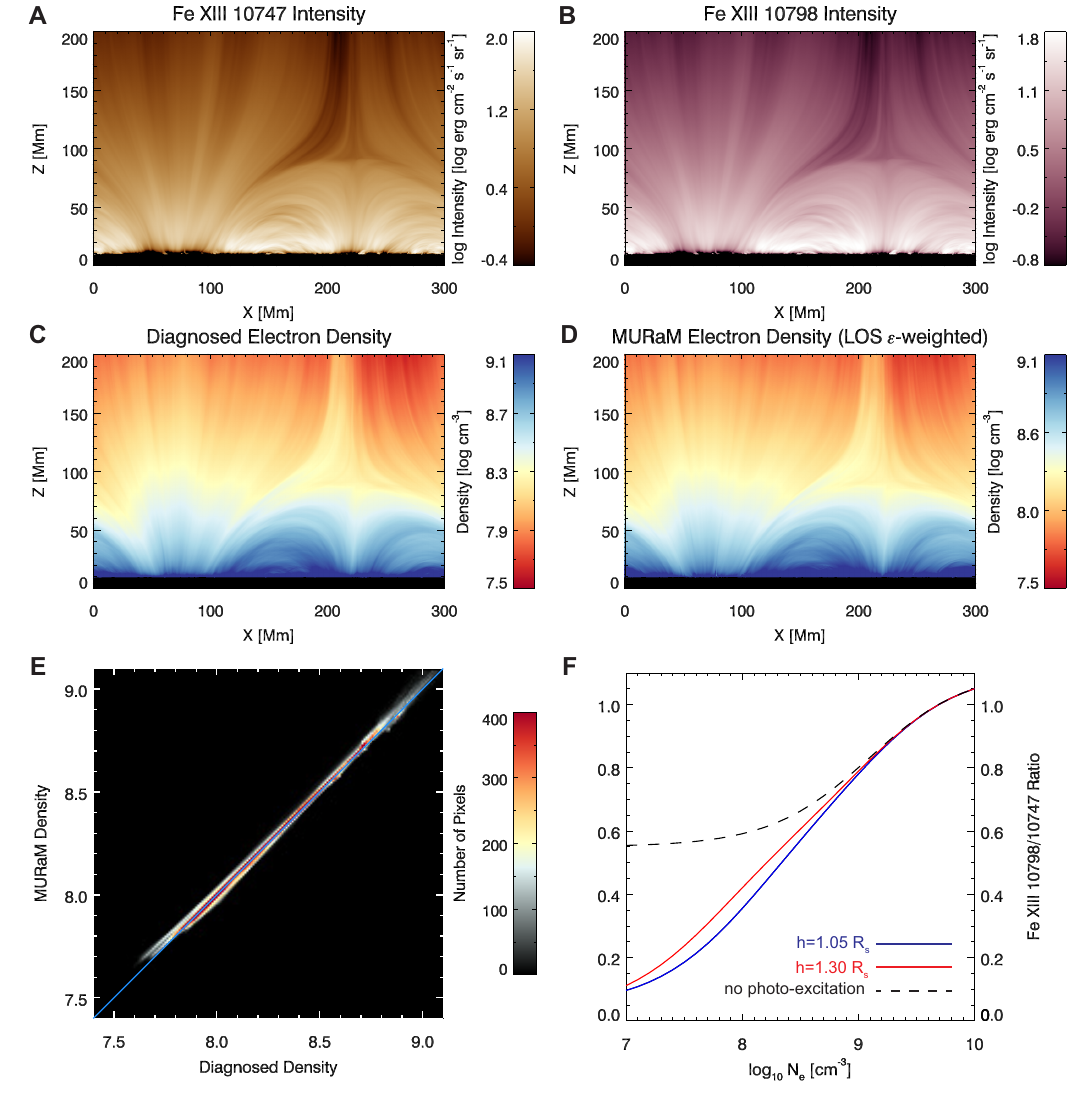}
    \caption{Diagnostics of electron density from synthetic Fe \sc{xiii}\rm{} line intensities. (A-B) Synthetic Fe \sc{xiii}\rm{} 10747 \AA\ and 10798 \AA\ intensities, respectively. Both collisional excitation and height-dependent photo-excitation are included in the synthesis. (C) Diagnosed electron density from the intensity ratio of the synthetic lines. (D) The LOS emissivity-weighted electron density from the MURaM simulation, representing the ground truth. (E) A 2D histogram showing the strong agreement between the diagnosed result and the ground truth. (F) Examples of the ratio-density curves with photo-excitation at the heights of 1.05 solar radii (blue solid curves) and 1.30 solar radii (red solid curves), respectively, and the curve without photo-excitation (black dashed curve).} 
    \label{fig:dens_int}
\end{figure}

To compare with the simulation ground truth, we computed the LOS emissivity-weighted electron density from the simulation data at each pixel as 
\begin{equation}\label{eq:ne}
    \overline{N_{e,0}}=\frac{\int{\varepsilon_i\cdot N_{e,i}\ dl}}{\int{\varepsilon_i\ dl}}
\end{equation}
where $N_{e,i}$ is the electron density at each point along the LOS. This provides a ground-truth density map directly comparable to the diagnostic result. Figure \ref{fig:dens_int} summarizes the comparison. Panels (A) and (B) show the synthetic Fe \sc{xiii}\rm{} 10747 \AA\ and 10798 \AA\ intensities, respectively. Panel (C) shows the diagnosed density from the line ratio, and panel (D) is the ground truth LOS emissivity-weighted electron density distribution from the simulation as calculated from Eq. \ref{eq:ne}. The 2D histogram comparing the diagnosed density with the ground truth in Figure \ref{fig:dens_int}(E) reveals close agreement between the diagnosed electron density and the ground truth. Panel (F) shows examples of the ratio-density curves with photo-excitation (blue and red solid curves) at different heights and without photo-excitation (black dashed curve) as calculated from \textsc{chianti}. In this study we adopt a viewing geometry in which the majority of the structures are oriented close to the POS and LOS density variations are moderate. Under these conditions, the LOS emissivity-weighted density serves as a reasonable representation of the density sampled by the line-ratio diagnostic, allowing for a direct comparison between the diagnosed and ground-truth density maps.

\subsection{Calculation of Wave Propagation Direction}
The wave propagation direction is inferred through cross-correlation of the Doppler velocity time series. As reported in previous work using CoMP/UCoMP observations \citep[e.g.,][]{2009ApJ...697.1384T,2015NatCo...6.7813M,2024Sci...386...76Y}, coherent propagating Doppler velocity fluctuations trace the travel direction of the transverse waves. For each pixel in the field-of-view (FOV), we perform cross-correlation between its Doppler velocity time series and those of nearby pixels to obtain a coherence map. Regions with high coherence appear as elongated lobes that indicate the local propagation trajectory of the waves. We estimate the propagation angle at each pixel by linear fitting the ridge of the elongated lobe. Repeating this step for all pixels in the FOV yields a 2D map of the wave propagation direction.

Since the transverse waves propagate along the magnetic field lines, the measured wave propagation direction should match the magnetic field direction projected onto the POS. To compare with simulation ground-truth, we computed the LOS emissivity-weighted POS field direction from the simulation cube as
\begin{equation}\label{eq:bangle_muram}
    \overline{\Psi_{\text{POS},0}}=\frac{\int{\varepsilon_i\cdot \Psi_{\text{POS},i}\ dl}}{\int{\varepsilon_i\ dl}}
\end{equation}
where $\Psi_{\text{POS},i}$ is the POS magnetic field direction at each point along the LOS, and $\varepsilon$ is the emissivity. We note that this choice avoids the $180^\circ$ ambiguity issue that arises if one first computes emissivity-weighted components (i.e., $B_\text{x}$ and $B_\text{z}$) of vector magnetic field and then derives the POS direction. In that case, along a LOS, two $B_\text{x}$ that are antiparallel (e.g., $+90^\circ$ and $-90^\circ$) would partially cancel when averaged, leading to an incorrect direction. By calculating the weighted field direction itself at each POS location after enforcing directional continuity, we ensure that such ambiguity does not distort the result. In principle, the alternative definition based on emissivity-weighted $B_\text{x}$ and $B_\text{z}$ components could also be used if the $180^\circ$-ambiguity can be reliably resolved along the LOS, but this is beyond the current scope.

\begin{figure}
    \centering
    \includegraphics[width=0.8\linewidth]{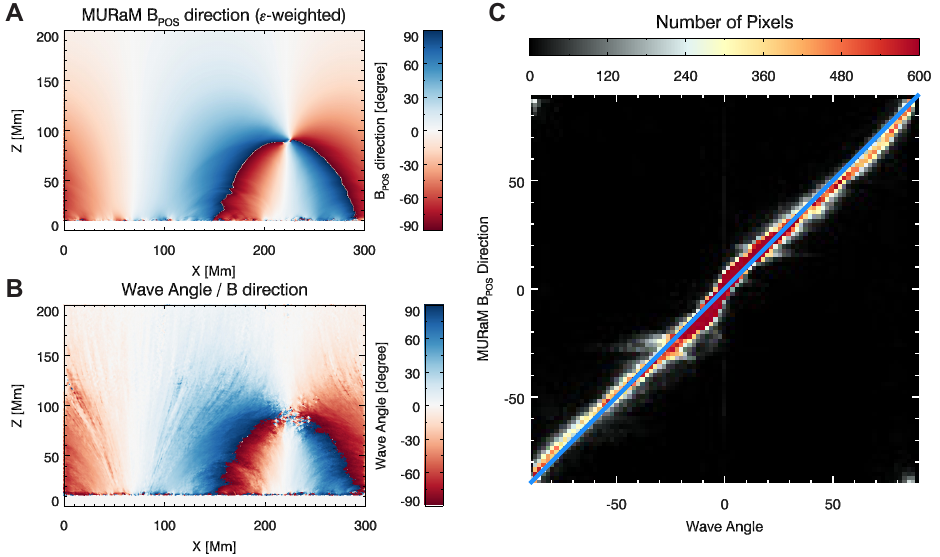}
    \caption{The calculation of wave propagation angle (magnetic field direction). (A) The ground-truth of the magnetic field direction: LOS emissivity-weighted average of the POS direction of magnetic field in the simulation. The reference direction is the vertical (radial) direction. The angle between the local magnetic field direction and the reference direction is defined as negative for clockwise rotation from the reference direction, and positive otherwise. (B) Calculated wave propagation angle using the wave-tracking procedure, which traces the POS magnetic field direction. The sign convention is the same as in panel A. (C) A 2D histogram showing strong agreement between the calculated wave angle and the ground truth magnetic field direction.}
    \label{fig:direction}
\end{figure}

Figure \ref{fig:direction}(A) and (B) are the measured wave propagation direction and the LOS emissivity-weighted POS magnetic field direction from MURaM simulation, respectively. The angles are measured with respect to the local radial (vertical in this case) direction, and are defined to be positive if the direction is counterclockwise to the local radial, and negative otherwise. A 2D histogram in Figure \ref{fig:direction}(C) comparing the two quantities reveals a strong correspondence between the two, indicating that the wave propagation direction obtained from the wave-tracking method recovers the POS magnetic-field direction, and most likely represents an emissivity-weighted LOS average.

\begin{figure}
    \centering
    \includegraphics[width=0.8\linewidth]{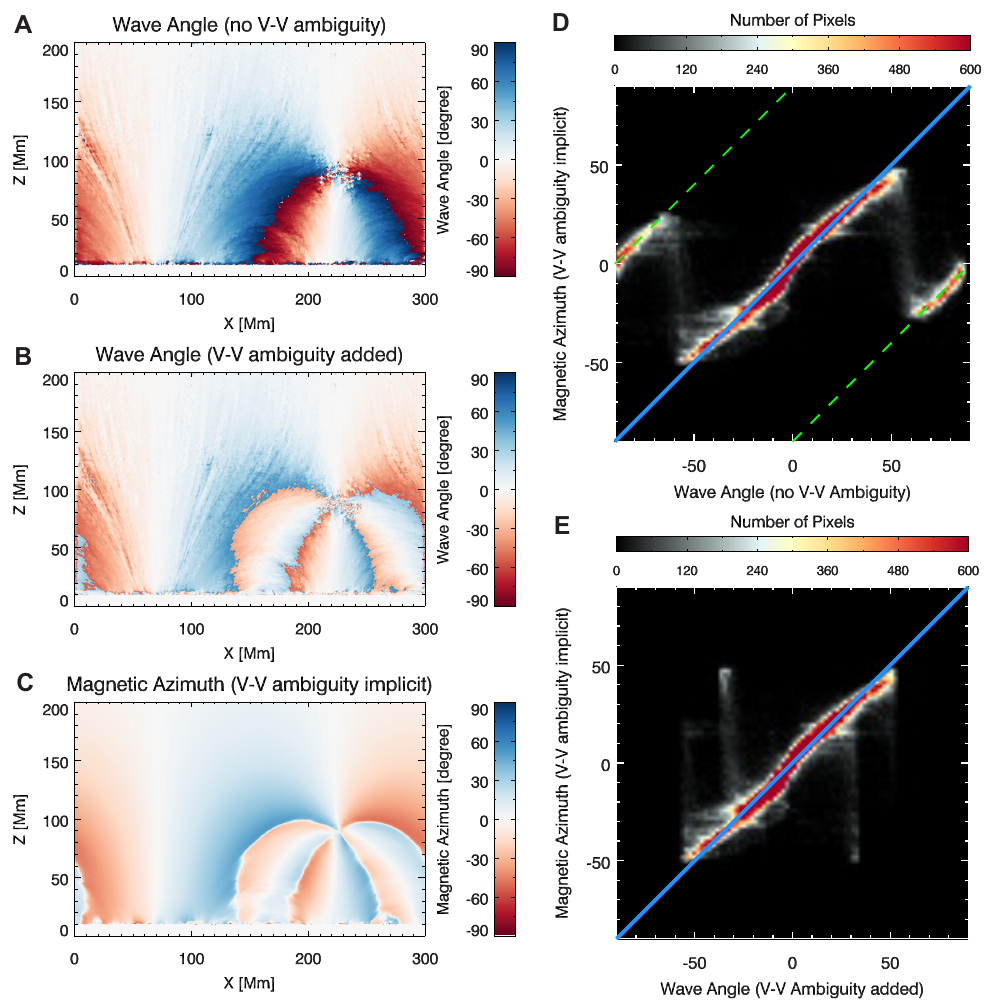}
    \caption{Comparison between the calculated wave propagation angle and magnetic azimuth. (A) Same as Fig. \ref{fig:direction}(B). This measurement is free of the Van Vleck ambiguity. (B) Similar to (A), but with Van Vleck ambiguity explicitly imposed to the calculated angles (angles $>54.7^{\circ}$ shifted by $-90^{\circ}$, angles $<54.7^{\circ}$ shifted by $+90^{\circ}$). (C) Magnetic azimuth derived from synthetic Stokes Q and U signals. The Van Vleck ambiguity is implicit in the azimuth. (D) A 2D histogram comparing the original wave propagation angle with the magnetic azimuth. In addition to the 1:1 ridge (blue line), two secondary ridges offset by $\pm 90^{\circ}$ arise from the Van Vleck ambiguity, which causes a $90^{\circ}$ rotation of the linear polarization when the magnetic field inclination exceeds the Van Vleck angle ($\sim 54.7^{\circ}$). (E) Similar to (D), but using the wave angle with Van Vleck ambiguity explicitly imposed. The two parameters now show close agreement.}
    \label{fig:azimuth}
\end{figure}

Another approach to diagnosing the POS magnetic field direction is to use the magnetic azimuth derived from linear polarization (Eq. \ref{eq:azimuth}).  Figure \ref{fig:azimuth}(C) shows the magnetic azimuth calculated from the synthesized Stokes Q and U signals. For Fe \sc{xiii}\rm{} 10747 \AA\ and 10798 \AA, which lie in the saturated Hanle regime, their linear polarization is subject to the Van Vleck ambiguity, whereby the polarization vector rotates by 90$^\circ$ when the field inclination relative to the local radial exceeds the Van Vleck angle of $\sim54.7^\circ$ \citep[e.g.,][]{1925PNAS...11..612V,1972SoPh...23..103H,2017ApJ...838...69L}. Consequently, in Figure \ref{fig:azimuth}(C), the regions where the true magnetic field angle is $>54.7^\circ$ or $<-54.7^\circ$ appear flipped by 90$^\circ$. 

By contrast, the wave propagation direction derived from wave tracking (Figure \ref{fig:azimuth}(A)) is free of the Van Vleck ambiguity and traces the actual projected POS field direction. The 2D histogram comparing the wave propagation direction with magnetic azimuth (Fig. \ref{fig:azimuth}(D)) exhibits the expected 1:1 ridge along the line of equality (blue solid line), together with two parallel ridges offset by $\pm 90^\circ$ (green dashed lines), characteristic of the Van Vleck ambiguity. 

To enable a direct comparison with the magnetic azimuth, in which the Van Vleck ambiguity is inherent, we explicitly imposed this ambiguity on the wave propagation directions: if $\Psi>54.7^\circ$ we set $\Psi\rightarrow\Psi-90^\circ$; if $\Psi<-54.7^\circ$ we set $\Psi\rightarrow\Psi+90^\circ$; otherwise $\Psi$ remain unchanged. It should be noted that $\Psi$ here does not represent the true field orientation but POS direction. In this work, we selected a viewing angle such that the majority of the magnetic structures are aligned with the POS, making this approximation reasonable. Through this process, we obtained a new wave propagation direction map with the Van Vleck ambiguity imposed manually, as shown in Fig. \ref{fig:azimuth}(B). The resulting map closely matches the distribution of the magnetic azimuth (Fig. \ref{fig:azimuth}(E)).

\subsection{Calculation of Wave Phase Speed and Magnetic Field Strength}\label{sec:wavetracking}

\begin{figure}
    \centering
    \includegraphics[width=0.9\linewidth]{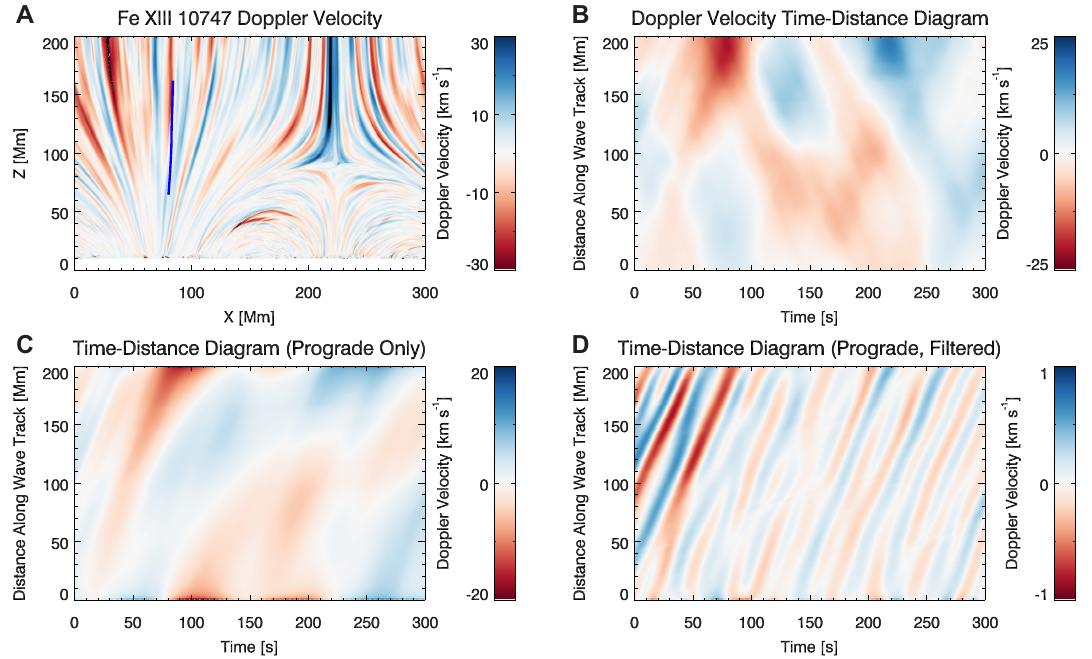}
    \caption{Calculation of wave phase speed. (A) One snapshot of the Fe \sc{xiii}\rm{} Doppler velocity image from the simulation. The blue line marks a representative long wave path. (B) Unfiltered Doppler velocity time-distance diagram derived along the wave path in panel A, showing signatures of counter-propagating waves components. (C) Doppler velocity time-distance diagram after isolating the prograde wave component. (D) Filtered version of panel C using the selected filtering frequency. Distinct inclined Doppler velocity ridges representing the propagating waves are clearly visible.}
    \label{fig:velo}
\end{figure}

With the wave propagation direction determined, we can estimate the wave phase speed. For each pixel in the FOV, we construct a fixed-length wave path based on the local wave angle; an example is shown as the blue track in Figure \ref{fig:velo}(A). Sampling the Doppler velocity along this path as a function of time produces a Doppler velocity time-distance diagram (Fig. \ref{fig:velo}(B)) that typically shows two counter-propagating branches of waves (prograde and retrograde). Following previous work, we isolate the prograde component to form the prograde-only time-distance diagram (Fig. \ref{fig:velo}(C)). The inclined ridges in this diagram are characteristic of propagating waves, and their slope is directly related to the phase speed. Practically, to compute the phase speed , we cross correlate the Doppler velocity time series at the path center with those at other positions to obtain the time lag as a function of distance along the wave path, then perform a linear fitting to the lag-distance relation to calculate the phase speed. Because the unfiltered time-distance pattern (Fig. \ref{fig:velo}(C)) from the original Doppler velocity is not strongly periodic, we bandpass the Doppler signal prior to cross-correlation to enhance coherence. In CoMP and UCoMP observations, a filtering frequency of 3.5 mHz is widely adopted, consistent with the observed power excess near this frequency, which is associated with the 5-minute coronal waves generated by p-modes \citep[][]{2019NatAs...3..223M}. In our MURaM simulation, however, the power spectrum lacks a distinct peak, and the power is broadly distributed across frequencies. This likely reflects a combination of factors related to the simulation setup. In particular, the finite spatial resolution results in convective granulation scales that are larger than solar (by approximately a factor of two), modifying the convective power spectrum and impacting the nature of p-mode excitation. In addition, the limited domain size provides an effective acoustic cavity that differs from that of the Sun. Given the short duration used in this work (a few hundred seconds), we adopt a higher filtering frequency of 0.05 Hz to ensure enough multiple wave cycles fit within the analysis window, improving the robustness of the cross-correlation. Figure \ref{fig:velo}(D) is the filtered Doppler velocity time-distance diagram. As shown later, the inferred phase speeds are not sensitive to the precise choice of filtering frequency provided it is chosen consistently with  the analysis scales and parameter constraints introduced below (see the Appendix \ref{app}).

\begin{figure}
    \centering
    \includegraphics[width=0.4\linewidth]{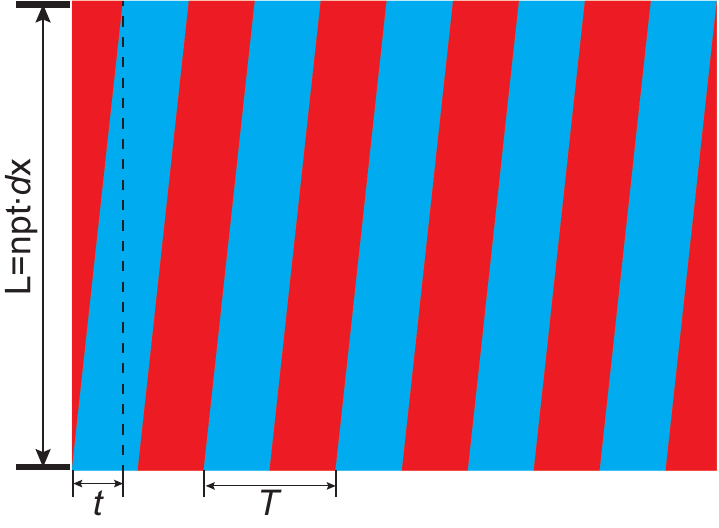}
    \caption{A schematic illustration of the parameters involved in accurate phase speed estimation from the Doppler velocity time-distance diagram. $L=\text{npt}\cdot d\text{x}$ is the length of the wave path, where npt is the number of pixels along the wave path and $d\text{x}$ is the spatial pixel size; $t$ is the travel time from one end to the other end of the Doppler velocity ridge, and $T$ is the filtering period which is directly related to the chosen filtering frequency $f$.}
    \label{fig:cartoon}
\end{figure}

Accurate phase speed estimation depends on spatial and temporal sampling and other parameters. As illustrated in Figure \ref{fig:cartoon}, to obtain accurate phase speed (i.e., the slope of the inclined Doppler velocity pattern), the Doppler velocity time series sampled across different positions along the wave path must exhibit measurable phase lags; equivalently, the ridge in the time-distance diagram should not approach a vertical orientation. In the near-vertical limit (very large slopes), the series become nearly synchronous and the cross-correlation result is poorly constrained, leading to less accurate phase speed estimations. In practice, as shown in Figure \ref{fig:cartoon}, if we let $t$ be the travel time from one end to the other end of the Doppler velocity ridge, and let $T=1/f$ be the filtering period (where $f$ is the filtering frequency), the ratio $t/T$ should be larger than a critical value to ensure the ridge is not in the near-vertical limit. If $v_{\text{ph}}$ is the phase speed, the length of the wave path is $L=\text{npt}\cdot \text{d}x$, where $\text{npt}$ is the number of pixels of the wave path and $\text{d}x$ is the spatial resolution of the data, then $t=L/v_\text{ph}$. To ensure $\alpha=t/T\geq \alpha_0$ where $\alpha_0$ is the critical value, we require the following relationship:
\begin{equation}\label{eq: alpha}
    \alpha=\frac{t}{T}=\frac{\text{npt}\cdot\text{d}x\cdot f}{v_\text{ph}}\geq \alpha_0
\end{equation}
This dimensionless parameter $\alpha$ controls whether phase lags are resolvable at the chosen scales. 

\begin{figure}
    \centering
    \includegraphics[width=0.9\linewidth]{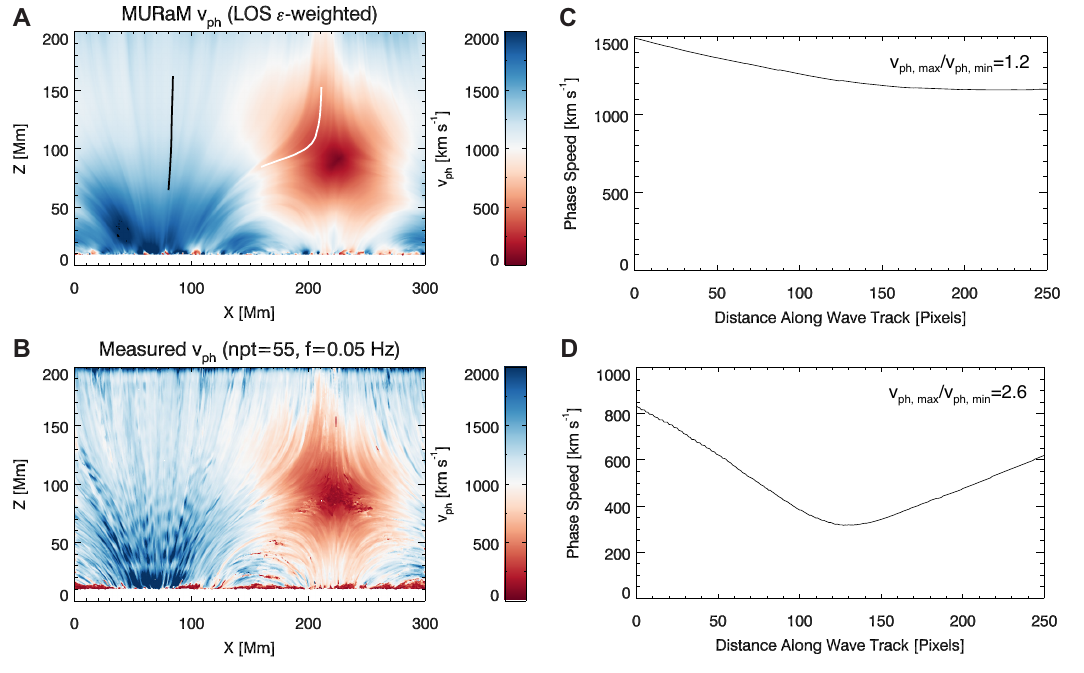}
    \caption{Calculation of wave phase speed and examples of phase-speed gradients in different regions. (A) LOS emissivity-weighted $v_\text{ph}$ from the simulation, representing the ground truth. The black line is an example wave path in a high-speed region, while the white curve shows a wave path in a low-speed region. (B) Phase speed map derived using the wave-tracking procedure with npt=55 and $f=0.05$ Hz. The result closely matches the ground truth. (C) Variation of phase speed along the black wave path in panel A. The ratio between the maximum and minimum phase speed along the path is around 1.2, indicating a small phase speed gradient. (D) Similar to (C) but for the white wave path in panel A. $v_\text{ph, max}/v_\text{ph, min}=2.6$ along this wave path, showing a strong phase speed gradient in this region.}
    \label{fig:vphmap}
\end{figure}

Figure \ref{fig:vphmap}(B) shows the calculated wave phase speed map using $f=0.05$ Hz, npt=55 pixels, and d$x$=0.384 Mm/pixel. To compare with the ground-truth phase speed distribution, we computed the LOS emissivity-weighted POS phase speed from the simulation as 
\begin{equation}
    \overline{v_{\text{POS}}}=\frac{\int{v_{\text{POS, i}}\cdot\varepsilon\ \text{d}l}}{\int{\varepsilon\ \text{d}l}}
\end{equation}
where $v_{\text{POS, i}}$ is the phase speed derived from POS magnetic field strength and density at each point along the LOS. Figure \ref{fig:vphmap}(A) is the ground truth phase speed map.

\begin{figure}
    \centering
    \includegraphics[width=0.9\linewidth]{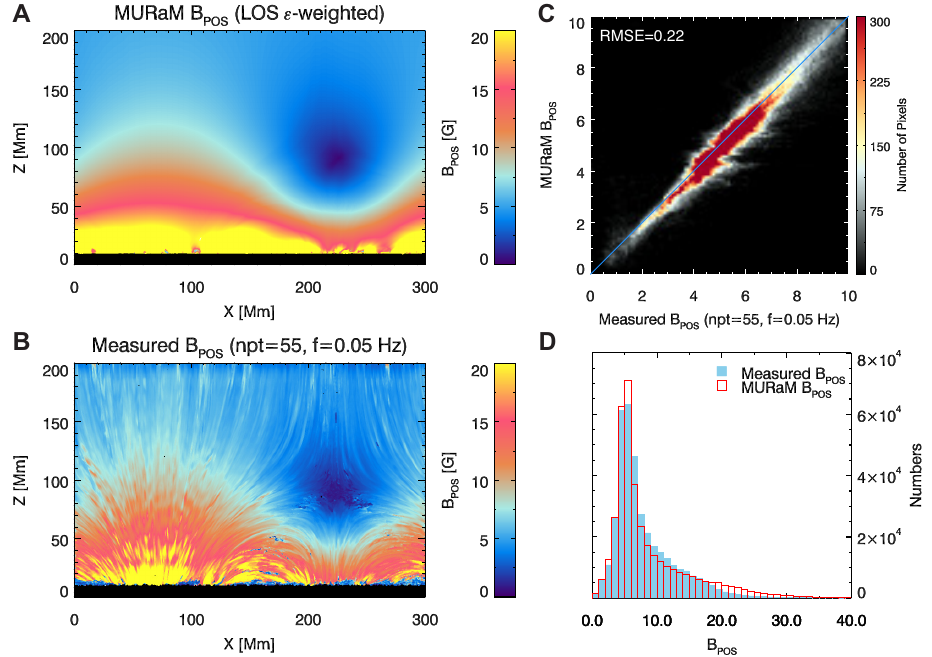}
    \caption{Estimation of magnetic field strength. (A) Ground-truth: LOS emissivity-weighted average of $B_\text{POS}$ from the simulation. (B) Measured magnetic field strength obtained by combining the density map in Fig. \ref{fig:dens_int}(C) and the phase speed map in Fig. \ref{fig:vphmap}(B) via Eq. 
    \ref{eq:vph}. (C) A 2D histogram comparing the measured and ground truth values. The calculated RMSE using Eq. \ref{eq:rmse} is 0.22, indicating close agreement. (D) Histograms showing the similar distribution of measured $B_\text{POS}$ and the ground-truth $B_\text{POS}$.}
    \label{fig:bmap}
\end{figure}

By combining the derived density map and phase speed map following Eq. \ref{eq:vph}, we obtain the magnetic field strength, as shown in Figure \ref{fig:bmap}(B). To provide a direct comparison with the simulation, we define the ground-truth magnetic field strength as 
\begin{equation}
    \overline{B_{\text{POS}}}=\frac{\int{B_{\text{POS, i}}\cdot\varepsilon\ \text{d}l}}{\int{\varepsilon\ \text{d}l}}
\end{equation}
where $B_{\text{POS, i}}$ is the POS magnetic field strength at each point along the LOS in the MURaM simulation. We note that this definition involves emissivity-weighted averaging of the magnetic field strength, rather than the vector magnetic field. This choice reflects the nature of the wave-based measurement: the inferred magnetic field is derived from the phase speed of transverse waves, which is a scalar quantity sensitive only to the magnitude of the magnetic field, not its direction. Therefore, the appropriate ground truth for comparison is the emissivity-weighted POS magnetic field strength rather than a weighted vector field. The ground truth map is shown in Fig. \ref{fig:bmap}(A). A 2D histogram in Fig. \ref{fig:bmap}(C) demonstrates that the measured magnetic field strength closely reproduce the LOS emissivity-weighted $B_\text{POS}$ from the simulation. To assess one-to-one correspondence, we calculate the root-mean-square of the error of $B_\text{obs}$ (the measured $B_\text{POS}$) relative to $B_\text{true}$ (the ground truth: LOS $\varepsilon$-weighted $B_\text{POS}$ from MURaM simulation) using
\begin{equation}\label{eq:rmse}
    \text{RMSE}=\sqrt{\frac{1}{N}\sum_i^N\left(\frac{B_{\mathrm{obs}, i}-B_{\mathrm{true}, i}}{B_{\mathrm{true}, i}}\right)^2}
\end{equation}
where $N$ is the number of data points, $B_{\mathrm{obs}, i}$ and $B_{\mathrm{true}, i}$ are the measured and ground-truth values, respectively. The histograms in Fig. \ref{fig:bmap}(D) compare the distributions of measured $B_\text{POS}$ and ground-truth $B_\text{POS}$, revealing large resemblance. These results confirm that the 2D coronal seismology technique, as applied to CoMP and UCoMP data, can provide accurate measurement of the coronal magnetic field, and the measured results represent a LOS emissivity-weighted average of the true POS magnetic field.

\section{Results and discussions}\label{sec: discussion}

\begin{figure}
    \centering
    \includegraphics[width=0.99\linewidth]{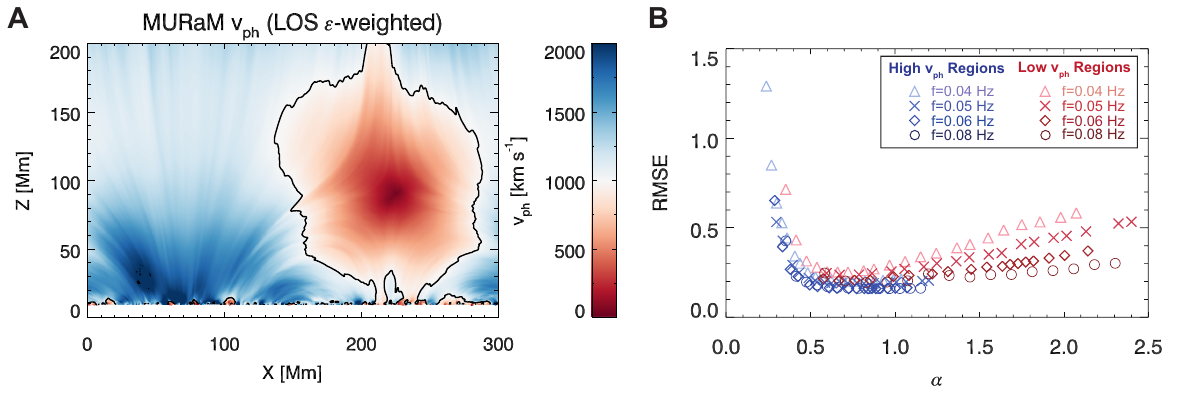}
    \caption{Variation of RMSE with varying $\alpha$. (A) Similar to Fig. \ref{fig:vphmap}(A), the black contour marks the location where $v_\text{ph}=1000\ \text{km}\ \text{s}^{-1}$. The region inside the contour (red color) is defined as low-speed region, while the surrounding area represents the high-speed region. (B) Dependence of RMSE on $\alpha$ and $f$, for both low $v_\text{ph}$ region and high $v_\text{ph}$ region. RMSE is very large with a small $\alpha$ value, which corresponds to near-vertical Doppler velocity ridges in the time-distance diagram. RMSE decreases rapidly as $\alpha$ increases, reaching a stable minimum around $\alpha\approx0.5$. This value represents the optimal lower threshold for accurate and robust performance. For $\alpha>0.5$, the results become largely insensitive to $\alpha$, especially in high-speed regions, while in low-speed regions excessively large $\alpha$ values lead to increased deviations from the ground-truth and a slight rise in RMSE.}
    \label{fig:rmse}
\end{figure}

Figures \ref{fig:direction} and \ref{fig:bmap} show that the 2D coronal seismology technique can accurately recover the emissivity-weighted magnetic-field strength and direction from the synthesized observables. To assess the robustness of the method, we performed a parameter-space analysis varying the wave-path length (npt) and filtering frequency ($f$). The results, summarized in Figure \ref{fig:rmse}, demonstrate that the diagnostic accuracy depends on the dimensionless parameter $\alpha$ as defined in Eq. \ref{eq: alpha}. To account for the broad range of phase speeds in the model, we separated the field of view into high-speed ($>$1000 km s$^{-1}$) and low-speed ($<$1000 km s$^{-1}$) regions and computed the RMSE between the inferred and ground-truth magnetic fields for each group. As shown in Figure \ref{fig:rmse}, different symbols denote results based on different filtering frequencies; blue symbols correspond to high-speed regions and red to low-speed regions. The black contours in Figure \ref{fig:rmse}(A) mark locations with phase speed of 1000 km s$^{-1}$. From Fig. \ref{fig:rmse}(B), RMSE decreases rapidly with increasing $\alpha$ and becomes largely insensitive once $\alpha\gtrsim 0.5$ in the high-speed regime, indicating that $\alpha\approx 0.5$ is a practical lower limit for robust diagnostics. For $\alpha$ below this threshold, time-distance ridges become nearly vertical and the phase lags are poorly constrained, leading to large uncertainties.

At the same time, $\alpha$ is not “the larger the better”. In the low-speed regions, excessively large $\alpha$ implies overly long wave paths that average across regions with strong phase-speed gradients, which raises the RMSE again. For example, in Fig. \ref{fig:vphmap}(A), the black and white curves are two representative wave paths in the high and low phase speed region, respectively. The phase speed variation along the two wave paths (Fig. \ref{fig:vphmap}(C) and (D)) show that the low-speed regions exhibits much stronger phase speed gradients than the high-speed region. In such cases, the averaging effect when calculating the phase speed from time-distance diagram will be more severe for the low-speed regions, leading to larger uncertainties of the measurement. This implies that optimal performance requires moderate $\alpha$ values slightly above 0.5, balancing between resolvable phase lags and minimal path averaging. The detailed parameter exploration and examples of magnetic field maps for different (npt, $f$) combinations are provided in Appendix \ref{app}. We further tested the sensitivity of our results to spatial resolution by degrading the synthetic observations, which are not presented here. While the reduced resolution leads to larger RMSE values due to fewer available data points within the field of view, the overall trends and main conclusions remain unchanged.

For observational context, typical CoMP settings use a spatial pixel size of $\sim 3.15\ \text{Mm}/\text{pixel}$ and a filtering frequency of 3.5 mHz, and the phase speed in the observed corona falls mostly below 700 km s$^{-1}$ \citep[e.g.,][]{2020ScChE..63.2357Y,2020Sci...369..694Y}. For UCoMP (dx $\approx$ 2.2 Mm/pixel) with the same npt and $f$ \citep[][]{2024Sci...386...76Y}, $\alpha\approx 0.34$, somewhat below the threshold, suggesting that slightly longer wave paths (e.g., npt$\approx45-55$) may improve accuracy, particularly in high-speed regions. We further propose an adaptive strategy: first estimate the spatial distribution of phase speed from an initial wave-tracking run, then adjust npt locally to maintain $\alpha\approx 0.5$. This approach yields consistent and robust diagnostic performance across the FOV. Nevertheless, the nearly flat RMSE behavior for $\alpha>0.5$ indicates that the technique is generally robust provided this criterion is satisfied.

From Figs. \ref{fig:dens_int}(C), \ref{fig:vphmap}(B) and \ref{fig:bmap}(B), filamentary structures are evident in the density, phase speed and magnetic field maps. The inferred density shows very similar features to the ground truth (Fig. \ref{fig:dens_int}(C) vs. Fig. \ref{fig:dens_int}(D)), and both the inferred and ground-truth phase speeds exhibit filamentary patterns, although the inferred one shows some differences \ref{fig:vphmap}(A) vs. Fig. \ref{fig:vphmap}(B)). In the simulation, the ground-truth phase speed is derived from the model magnetic field and density. The model magnetic field is relatively smooth because it is dominated by large-scale flux systems, whereas the density shows small-scale variations. As a result, the ground-truth phase speed naturally inherits these filamentary features from the density distribution (Fig. \ref{fig:vphmap}(A)). In contrast, comparing Fig. \ref{fig:bmap}(A) and (B), while the ground-truth magnetic field is smooth, the inferred magnetic field appears highly structured. This is mainly because the POS magnetic field strength at each slice along the LOS is smooth, and its variation along the LOS is small, thus the weighted magnetic field (the ground truth) is also smooth. However, for the inferred magnetic field, because the seismological inversion involves $B\propto v_\text{ph}\sqrt{\rho}$, any localized fluctuations in the derived $v_\text{ph}$ or $\rho$, that deviate from the ground truth will contribute to the observed fine-scale structures in the inferred magnetic field map (Fig. \ref{fig:bmap}(B)). We also performed a simple test based on a simple 2D numerical model showing that, because the phase speed is obtained by cross-correlating Doppler velocity time series of limited duration, the initial phase at the beginning of each time series can slightly shift the cross-correlation peak due to the finite temporal window. This introduces small differences in the recovered $v_\text{ph}$, and hence $B$, across neighboring field lines, partly contributing to the filamentary features. 

A second feature is the periodic variations of measured phase speed (Fig. \ref{fig:vphmap}(B)) and magnetic field strength (Fig. \ref{fig:bmap}(B)) along field lines, most apparent in open field regions. In the simulation, waves reflect at the top boundary, so prograde and retrograde wave components have comparable power and can interfere. Although the wave-tracking procedure isolates the prograde component \citep[e.g., ][]{2020Sci...369..694Y, 2024Sci...386...76Y}, if the two components have similar amplitudes and wave power, strong interference at specific locations can complicate the time-distance ridge and result in a quasi-periodic pattern along the path. A supplemental simulation with greatly reduced top-boundary reflections shows that, when retrograde power is minimal, the measured phase speed shows almost no periodic variations. This supports the view that the periodic patterns in our main simulation arise from interference between counter-propagating waves. In CoMP and UCoMP observations, however, the prograde wave components typically dominates in amplitude and power \citep[e.g.,][]{2009ApJ...697.1384T, 2024Sci...386...76Y}, so the interference effect appears to be minor and no obvious periodic patterns were identified. A more detailed investigation of the origin and physical implications of the fine-scale and periodic features identified here will be presented in our forthcoming work.

\section{Summary and Future Perspectives}

The recently developed 2D coronal seismology has been successfully applied to CoMP and UCoMP spectral imaging observations, enabling routine measurements of the global coronal magnetic field. To validate the accuracy and robustness of this technique, we carried out a forward modeling analysis based on a MURaM simulation. By applying the 2D coronal seismology to synthetic observables (Fe \sc{xiii}\rm{} 10747 \AA\ and 10798 \AA\ line intensity, Fe \sc{xiii}\rm{} 10747 \AA\ line Doppler velocity and Fe \sc{xiii}\rm{} 10747 \AA\ linear polarization signals) from the simulation, we retrieved the magnetic field direction and strength. The diagnosed results show strong agreement with the simulation ground-truth, confirming the accuracy of the technique in measuring both the direction and strength of coronal magnetic field. The forward modeling result also suggests that the measured coronal magnetic field corresponds to a LOS emissivity-weighted average of the true magnetic field, consistent with previous analyses of UCoMP data \citep[][]{2024Sci...386...76Y}. This work is a first step toward validating the methodology. In real observations, uncertainties in the density diagnostics, phase-speed estimation, and measurement errors will all contribute to the uncertainty in the inferred magnetic field. 

To evaluate the performance of this technique under different settings, we established a relations linking several key parameters in the wave-tracking procedure and defined a dimensionless parameter $\alpha$ to constrain the required parameter setup (filtering frequency, number of length of the wave track, spatial resolution and maximum phase speed in the target region) during wave-tracking procedure. A parameter-space analysis (for detail see the Appendix \ref{app}) varying the number of wave path length and filtering frequency show that the 2D coronal seismology can provide reliable measurements as long as the parameters settings used during wave tracking satisfies $\alpha>0.5$. The diagnostics generally remains stable for larger $\alpha$, but we recommend keeping $\alpha$ close to 0.5 for optimal performance. This analysis provides guidance for selecting parameters in future observations from CoMP/UCoMP-like instruments.

It is important to note that some patterns in the reconstructed magnetic field maps (e.g., fine-scale and periodic patterns) may arise from the wave-tracking procedure and may not reflect the actual structures in the magnetic field. Future work should explore these effects in more detail to further refine the technique. In addition, our simulation employs a relatively short LOS integration path ($\sim$ 49 Mm), and the viewing angle we selected makes the majority of the structures aligning with the POS so there are less complicated coronal structures along the LOS. Testing the method with more complex simulations (e.g., simulations with more complicated LOS structures) will help determine the robustness under varied coronal conditions. Observational noise was also not included in the present study; incorporating noise in synthetic data will be an important next step toward understanding and improving the reliability of 2D coronal seismology under real observational conditions.

\begin{acknowledgments}
This material is based upon work supported
by the NSF National Center for Atmospheric Research, which is a major facility
sponsored by the U.S. National Science Foundation under Cooperative Agreement No.
1852977. We would like to acknowledge high-performance computing support from the Derecho system (doi:10.5065/qx9a-pg09) provided by the NSF National Center for Atmospheric Research (NCAR), sponsored by the National Science Foundation. Z.Y. acknowledges Dr. Giulio Del Zanna for helpful discussion on coronal infrared line synthesis. 

\end{acknowledgments}

\software{CHIANTI, FORWARD}

\appendix

\section{Parameter-space analysis and optimal $\alpha$ condition}\label{app}

As noted in Sect. \ref{sec:wavetracking}, accurate phase-speed estimation with wave tracking depends on the number of pixels along the wave path (npt), the spatial resolution (d$x$), the filtering frequency ($f$), and the actual phase speed ($v_\text{ph}$). These parameter in general should satisfy the relationship defined by Eq. \ref{eq: alpha}. Accurate phase-speed estimation requires $\alpha$ to exceed a critical value ($\alpha_0$), ensuring that measurable phase lags exist between time series at different positions along the path. To determine the optimal parameter settings for reliable and robust diagnostics, we carried out a parameter-space analysis by varying the number of pixels along each wave path (npt) and the filtering frequency ($f$). It is to be noted that we did not change d$x$ and $v_\text{ph}$, as the spatial resolution and the phase speed at every point in the simulation are already predefined. For real observations, the spatial resolution is also a known parameter, while the phase speed in the regions of interest can be estimated by an initial run of the wave tracking.

We performed the 2D coronal seismology analysis for synthetic Doppler velocity data filtered at $f=$0.04 Hz, 0.05 Hz, 0.06 Hz and 0.08 Hz, respectively. For filtered data under each filtering frequency, we also varied the number of pixels of the wave path (npt). In principle, $v_\text{ph}$ is not a single fixed number across the FOV. However, in practice, we could select a representative $v_\text{ph}$ value that is higher than the phase speed of most regions in the FOV. In this case, we can make sure that the derived $\alpha$ in Eq. \ref{eq: alpha} is fixed as a lower limit. With $\text{d}x$ and $v_\text{ph}$ fixed, each combination of npt and $f$ corresponds to a value $\alpha$, and for each combination of npt and $f$, we can calculate the resulted RMSE of the measured $B_\text{POS}$ relative to the ground truth. Figure \ref{fig:rmse}(B) shows the distribution of RMSE as a function of $\alpha$. As mentioned in Sect. \ref{sec: discussion}, we split the FOV into regions with low phase speed and high phase speed. The values of $\alpha$ for the two regions are determined as $\alpha_\text{low}=384\cdot\text{npt}\cdot f/1000$ and $\alpha_\text{high}=384\cdot\text{npt}\cdot f/2000$, respectively, where the spatial pixel size $\text{d}x=384\ $km, $v_\text{ph, low}=1000$ km s$^{-1}$ for the low speed regions and $v_\text{ph, high}=2000$ km s$^{-1}$ for the high speed regions. 

\begin{figure}
    \centering
    \includegraphics[width=1.0\linewidth]{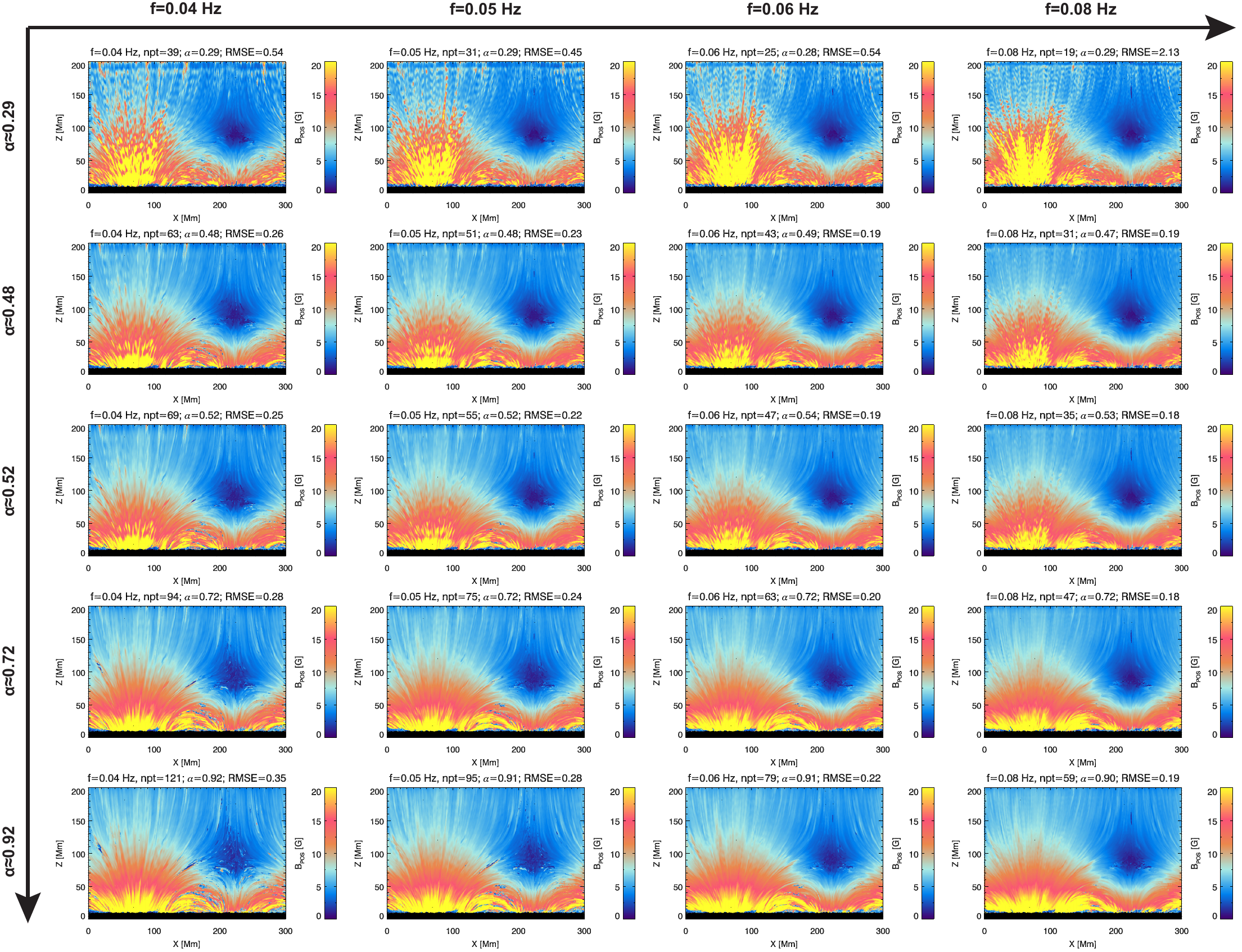}
    \caption{Measured magnetic field strength maps for different $f$, npt and $\alpha$ as in Figure \ref{fig:rmse}. Each row represents the results under different $\alpha$, while each column corresponds to different $f$. For each $f$, npt is chosen to achieve the indicated $\alpha$.}
    \label{fig:bmap_all}
\end{figure}

In Figure \ref{fig:bmap_all}, we show examples of the diagnosed magnetic field maps with different $f$ and npt (thereafter different $\alpha$). The calculated RMSE values are also provided in each panel. When npt is too small (small $\alpha$), the sampling along the wave path is insufficient; in high-speed regions, the Doppler velocity ridges in the time-distance diagram become nearly vertical, phase lags are poorly constrained, and the uncertainty on $v_\text{ph}$ becomes large. This explains the overestimated B in strong-field/high-speed areas in the first row of Figure \ref{fig:bmap_all}. As npt increases, it adds sampling points, improves the lag-distance fit, and the derived magnetic field strength approaches the ground truth. Once $\alpha$ exceeds about 0.5, the discrepancy between the inferred and ground-truth magnetic field is minimized (e.g., the third row of Fig. \ref{fig:bmap_all}). Beyond this point, further increasing npt brings little accuracy gain, but the magnetic field maps appear smoother because of longer path-averaging. This averaging effect may smooth the periodic variations of phase speed mentioned in Sect. \ref{sec: discussion}. In regions with strong phase speed gradient (e.g., low-speed/weak-magnetic-field-strength regions in Fig. \ref{fig:bmap_all}), very long paths bias the result through excessive averaging and increased discrepancy, sometimes even produce even noisier patterns in dark blue regions in Fig. \ref{fig:bmap_all}. Therefore, longer wave paths are not always preferable, especially for low-speed regions. As proposed in Sect.\ref{sec: discussion}, an adaptive approach by adjusting npt for different regions may improve performance.

Fig. \ref{fig:rmse}(B) also shows that, at the same $\alpha$, higher $f$ yields slightly lower RMSE and smoother magnetic field maps. For a time series of fixed duration, a higher filtering frequency leads to a higher number of oscillation cycles within the analysis window, improving cross-correlation precision. In practice, CoMP and UCoMP sequences (typically $>$ 1 hour) contain sufficient cycles for reliable cross-correlation.

\bibliography{sample701}{}
\bibliographystyle{aasjournalv7}

\end{CJK*}
\end{document}